\documentclass[manuscript]{aastex63}

\accepted{}
\submitjournal{ApJ}

\shorttitle{The QBO-type signals in the subsurface rotation rate residuals}
\shortauthors{Inceoglu et al.}

\begin{document}

\title{The QBO-type signals in the subsurface flow fields during solar cycles 23 and 24}

\correspondingauthor{Fadil Inceoglu}
\email{fadil.inceoglu@colorado.edu}

\author[0000-0003-4726-3994]{Fadil Inceoglu}
\affiliation{Cooperative Institute for Research in Environmental Sciences, University of Colorado Boulder, Boulder, CO, USA}
\affiliation{National Centers for Environmental Information, National Oceanographic and Atmospheric Administration, Boulder, CO, USA}


\author{Rachel Howe}
\affiliation{School of Physics and Astronomy, University of Birmingham, Edgbaston, Birmingham B15 2TT, UK}
\affiliation{Stellar Astrophysics Centre (SAC), Department of Physics and Astronomy, Aarhus University, Ny Munkegade 120, DK-8000 Aarhus C, Denmark}

\author{Paul T. M. Loto'aniu}
\affiliation{Cooperative Institute for Research in Environmental Sciences, University of Colorado Boulder, Boulder, CO, USA}
\affiliation{National Centers for Environmental Information, National Oceanographic and Atmospheric Administration, Boulder, CO, USA}




\begin{abstract}

We studied the presence and spatiotemporal evolution of the quasi-biennial oscillations (QBOs) in the rotation rate residuals at target depths of 0.90$R_{\odot}$, 0.95$R_{\odot}$, and 0.99$R_{\odot}$ and at low (0 -- 30$^{\circ}$), mid (30 -- 50$^{\circ}$), and high (50 -- 70$^{\circ}$) latitudinal bands. To achieve these objectives we used data from the Michelson Doppler Imager (MDI) on {\it the Solar and Heliospheric Observatory} ({\it SOHO}) and the Helioseismic and Magnetic Imager (HMI) on the {\it Solar Dynamics Observatory} ({\it SDO}), covering solar cycles 23 and 24, respectively. The results show that there are QBO-like signals in each latitudinal band and depth however they are affected by higher amplitude and longer-time scale variations. The QBO-like signals found in each target depth and latitudinal bands show different spatiotemporal evolution. The amplitudes of variations of the rotation rate residuals in the QBO timescale increase with increasing depth.

\end{abstract}

\keywords{QBO, rotation rate, differential rotation, wavelet}


\section{Introduction} \label{sec:intro}

As a magnetically active star, the Sun shows cyclic variations in its activity levels in timescales ranging from months to decades \citep{1939Obs....62..158G, Suess1980Raiocarbon, 1990SoPh..129..165P}. The most known among these cyclic activities is the Schwabe cycle \citep{1844AN.....21..233S}, where the observed sunspot numbers vary approximately with an 11-year cycle. In addition to these, the Sun also shows quasi-biennial oscillations (QBOs). Earlier studies suggested that the period of the QBO range from 1.5 to 4 years \citep{2009A&A...502..981V} as well as from 1 to 3 years \citep{10.1093/mnras/stz1653}. In the literature, the periodicities of the QBOs are generally suggested to range from 0.6 to 4 years, where there is a clear separation at 1.5 years pointing to two groups of variations below and above this value \citep{2014SSRv..186..359B}. The amplitude of the QBOs are in phase with the Schwabe cycle, attaining their highest amplitude during solar cycle maxima and become weaker during solar cycle minima. The QBOs are observed to behave differently in each solar hemisphere while showing the same intermittency \citep{2014SSRv..186..359B, 2019A&A...625A.117I}. Together with exhibiting signals distributed over all solar latitudes in the magnetic synoptic maps \citep{2012ApJ...749...27V}, they are also reported to be present in the high solar latitudes ($\ge 60^{\circ}$) using monthly values of the polar faculae data recorded between 1951 and 1998 \citep{2020MNRAS.494.4930D}.

The physical mechanisms that could explain the generation and spatiotemporal evolution of the QBOs include a secondary dynamo working in the subsurface shear layer at 0.95R$_{\odot}$ \citep{1998ApJ...509L..49B,2010ApJ...718L..19F}, a 180$^{\circ}$ shifting of the active longitudes \citep{2003A&A...405.1121B}, spatiotemporal fragmentation of radial profiles of rotation rates \citep{2013ApJ...765..100S}, instability of the tachocline magnetic Rossby waves \citep{2010ApJ...724L..95Z} and tachocline non-linear oscillations (TNOs), where periodically varying energy exchange takes place between the Rossby waves, differential rotation, and the present toroidal field \citep{2018ApJ...853..144D}. Recently, using a flux transport dynamo code, \citet{2019A&A...625A.117I} proposed that there are indications for the QBOs to be generated via interplay between the flow and magnetic fields, where the turbulent $\alpha$-mechanism working in the lower half of the solar convection zone generates poloidal fields from an existing toroidal field.

In this study, we study the presence of the QBOs in low (0 -- 30$^{\circ}$), mid (30 -- 50$^{\circ}$), and high (50 -- 70$^{\circ}$) solar latitudes in solar subsurface flows at the target depths of 0.90$R_{\odot}$, 0.95$R_{\odot}$, and 0.99$R_{\odot}$, respectively, using data from the Michelson Doppler Imager (MDI) on {\it the Solar and Heliospheric Observatory} ({\it SOHO}) and the Helioseismic and Magnetic Imager (HMI) on the {\it Solar Dynamics Observatory} ({\it SDO}), covering solar cycles 23 and 24. We first describe the data sets used in this study (Section~\ref{sec:data}). We analyze the data to decipher the spatiotemporal features of the QBOs in low, mid, and high solar latitudes and present the results in Section~\ref{sec:analyses_res}. Then we discuss the results and conclude in Section~\ref{sec:dis_conc}.

\section{Data} \label{sec:data}

\subsection{Subsurface Flow Fields from MDI/SOHO and HMI/SDO}

\begin{figure*}
\begin{center}
{\includegraphics[width=5.5in]{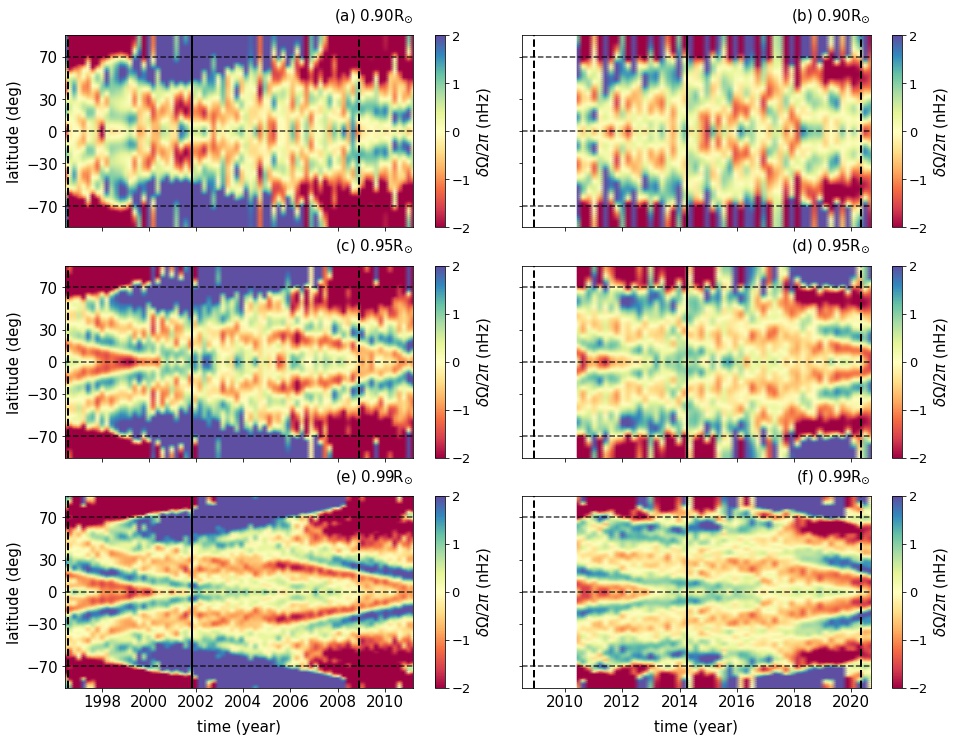}}
\caption{Panels (a), (c), and (e) show the rotation-rate residuals as a function of latitude at a target depths of 0.90$R_{\odot}$, 0.95$R_{\odot}$, and 0.99$R_{\odot}$, respectively, for solar cycle 23 from MDI/SOHO data. Panels (b), (d), and (f) show the same for solar cycle 24 from HMI/SDO data. The vertical solid-lines show solar cycle maxima, while vertical dashed-lines show starts and ends of the solar cycles.}
\label{fig:mdi_hmi_mag_flow}
\end{center}
\end{figure*}

The rotation rates are calculated based on regularized least squares (RLS) code using frequencies derived from MDI data. The MDI data consists of 74 non-overlapping 72 day observation periods starting in May 1996. To calculate the rotation rates between May 1996 and February 2011, 74 sets of rotational-splitting coefficients up to $l = 300$ from MDI were used. The rotation rate residuals are calculated by removing the temporal mean from the MDI data \citep[for detailed information about the methods, see][]{2009LRSP....6....1H, 2018ApJ...862L...5H}. It must be noted that the inversion data do not distinguish between the northern and southern solar hemispheres, therefore the values are flipped around the equator (Figures~\ref{fig:mdi_hmi_mag_flow}a, c and e). 

To calculate the rotation rates for the period covering solar cycle 24, we again employed the RLS code using frequencies derived from the HMI data, which consists of 53 non-overlapping 72 day observation periods starting in May 2010. To achieve this, we used 53 sets of rotational-splitting coefficients up to $l = 300$ from HMI covering the period from May 2010 to August 2020. The rotation rate residuals, similar to those from MDI, are calculated by removing the temporal mean from those inferred from the HMI data set ( Figures~\ref{fig:mdi_hmi_mag_flow}b, d, and f).

We decided not to splice the data sets from HMI and MDI as they only have an overlapping period of one year where the solar cycle is at its minimum. Using one year overlapping period over the solar cycle minimum might introduce systematic errors.

Interestingly, although the rotation rate residuals show similar features until 2 years before each solar cycle end, the rotation rate residuals at each depth 0.90$R_{\odot}$, 0.95$R_{\odot}$, and 0.99$R_{\odot}$ display opposite behavior after this time. Between 2006 and 2009, a pole-ward slower-than-average flow band forms around 50$^{\circ}$ latitude (Figure~\ref{fig:mdi_hmi_mag_flow}a, c, and e). Between May 2010 and September 2020 during solar cycle 24, on the other hand, the slower-than-average flow band forms around 60$^{\circ}$ latitude, and then it propagates equator-ward (Figure~\ref{fig:mdi_hmi_mag_flow}b, d, and f). The physical mechanism which might lead to the observed differences in rotation rate residuals is to be studied further and not within the scope of this study. The acceleration and deceleration compared to the average background flow during solar cycle 24 are slightly weaker compared with those in solar cycle 23.

\section{Analyses and Results} \label{sec:analyses_res}

\begin{figure*}
\begin{center}
{\includegraphics[width=4in]{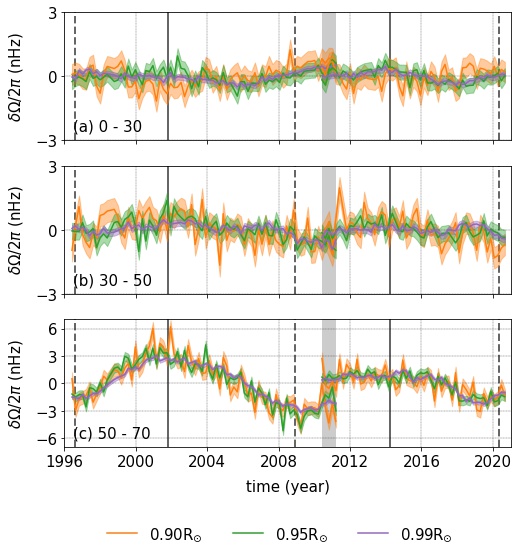}}
\caption{The rotation-rate residuals at target depths of 0.90$R_{\odot}$ (orange), 0.95$R_{\odot}$ (green), and 0.99$R_{\odot}$ (purple) for the latitudinal bands 0 -- 30$^{\circ}$ (a), 30 -- 50$^{\circ}$ (b), and 50 -- 70 $^{\circ}$ (c) covering the period between May 1996 -- November 2008 from MDI/SOHO and the period between May 2010 -- September 2020 from HMI/SDO. Color shaded areas show the error interval calculated for the flow fields. The vertical solid lines show the solar cycle maxima, while the dashed lines show the cycle minima. The vertical grey shaded are show the overlapping time period between two data sets.} 
\label{fig:mdi_hmi_mag_lats}
\end{center}
\end{figure*}

Before any further analysis, we first interpolated the MDI data set to have equal time increments of 72 days between the data points. There were two data points which were separated by $\sim$180 and  $\sim$108 days due to the known SOHO mission interruptions occurred in 1998\footnote{\url{https://www.esa.int/esapub/bulletin/bullet97/vandenbu.pdf}} and early 2019 \citep{2004SpWea...2.2004D}. Following that, we calculated the average rotation-rate residual values for the latitudinal bands 0 -- 30$^{\circ}$ (low), 30 -- 50$^{\circ}$ (mid), and 50 -- 70 $^{\circ}$ (high) latitudinal band. We need to note that since the rotation-rate residuals are flipped around the equator, we did not indicate any hemisphere in the relevant plot (Figure~\ref{fig:mdi_hmi_mag_lats}). It must also be noted that we did not remove the annual effects from the data as the cut-off frequency of a possible filter would be very close to the lower end of the QBO periods.

\subsection{The temporal evolution of rotation rate residuals}

The rotation rate residuals at each depth show higher frequency fluctuations than the 11-year cycle throughout solar cycle 23 at the low and mid latitudinal bands (Figures~\ref{fig:mdi_hmi_mag_lats}a and b). The rotation rate residuals for the high latitudinal band, on the other hand, exhibits variations in phase with the solar cycle. The acceleration compared to the mean background flow trend is seen at the onset of solar cycle 23, which then starts to gradually decelerate after around 2001 until 2006 (Figure~\ref{fig:mdi_hmi_mag_lats}c). Another interesting feature is that the amplitudes of higher frequency variations in the rotation rate residuals increase with increasing depth (Figure~\ref{fig:mdi_hmi_mag_lats}).

Similar to solar cycle 23, solar cycle 24 shows high-frequency variations (Figures~\ref{fig:mdi_hmi_mag_lats}). However, different than solar cycle 23, the high latitudinal band in solar cycle 24 does not exhibit in phase variations with the solar cycle. The rotation rate residuals show a very different behavior; the rotation rate residuals fluctuate around 1 nHz. After around 2017, the rotation rate residuals show slower-than-average values followed by a recovery trend by 2020 (Figure~\ref{fig:mdi_hmi_mag_lats}c). This behavior is related to the equator-ward slower-than-average flow band observed in the higher latitudes in solar cycle 24 (the right panel of Figure~\ref{fig:mdi_hmi_mag_flow}). Rotation rate residuals, similar to solar cycle 23, show increasingly higher amplitude variations as we go deeper layers into the Sun (Figure~\ref{fig:mdi_hmi_mag_lats}).

To investigate the presence of the QBOs in low, mid, and high latitudinal bands in the subsurface flows, we use continuous wavelet transformation (CWT) following \citet{1998BAMS...79...61T}. We must note that lag-1 autocorrelation coefficients, which are used to calculate the red noise spectrum, could not be calculated for some latitudinal bands due to numerical reasons such as insufficient length of data or strong trends in it. Therefore, the significances of the CWTs are calculated against white noise at significance levels of 0.1, 0.5, and 0.01. 

\begin{figure*}
\begin{center}
{\includegraphics[width=6in]{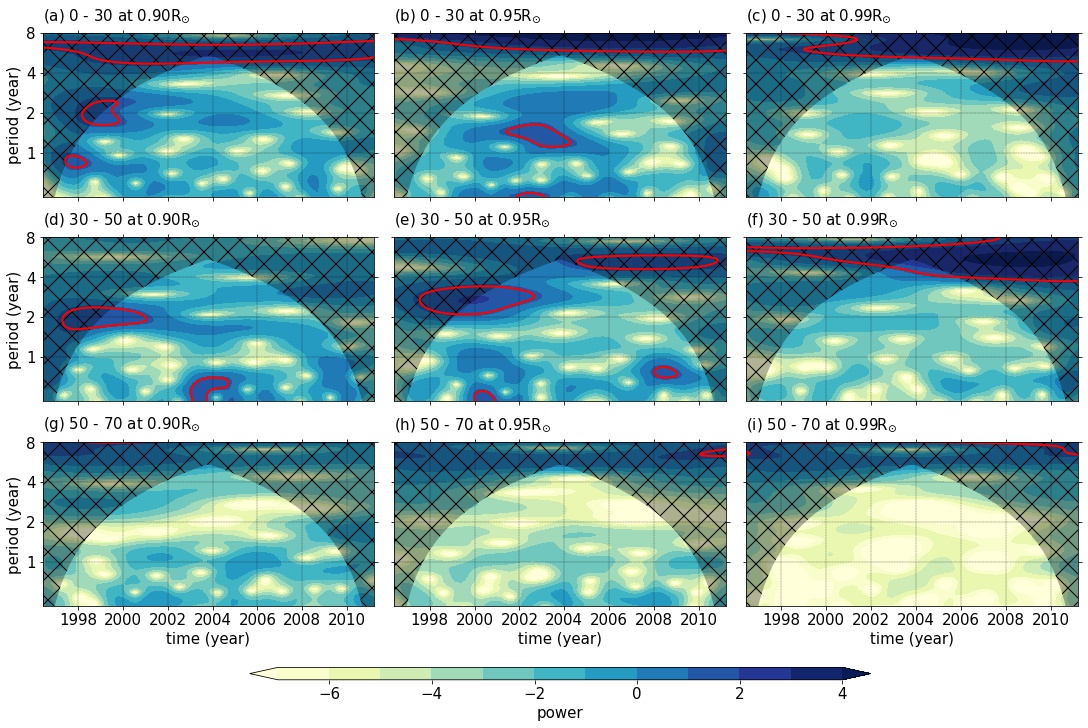}}
\caption{CWTs of the rotation rate residuals at 0.90$R_{\odot}$ (the left panel), 0.95$R_{\odot}$ (the middle panel), and 0.99$R_{\odot}$ (the right panel) for the low, mid, and high latitudinal bands for the period covering May 1996 -- November 2008 from MDI/SOHO. The red contours represent the significance level of 0.1, while the hatched area marks the cone of influence, where the CWTs might be influenced by the edge-effects.} 
\label{fig:mdi_flow_CWT}
\end{center}
\end{figure*}

During solar cycle 23, even though the rotation rate residuals exhibit QBO-like signals in general, only some of these signals are statistically significant at the 0.1 level. The rotation rate residuals at the target depth of 0.90$R_{\odot}$ at the low and mid latitudinal bands exhibit QBO-like signals during the rising phase of solar cycle 23 (the left panel of Figure~\ref{fig:mdi_flow_CWT}). At the depth of 0.95$R_{\odot}$, the low latitudinal band shows a statistically significant QBO-like signal between 2000 and 2006, while the mid latitudinal band at the same depth shows a QBO-like signal only on the onset of the solar cycle (the middle panel of Figure~\ref{fig:mdi_flow_CWT}). There are not any statistically significant QBO-like signals detected at the depth of 0.99$R_{\odot}$ (the right panel of Figure~\ref{fig:mdi_flow_CWT}).

\begin{figure*}
\begin{center}
{\includegraphics[width=6in]{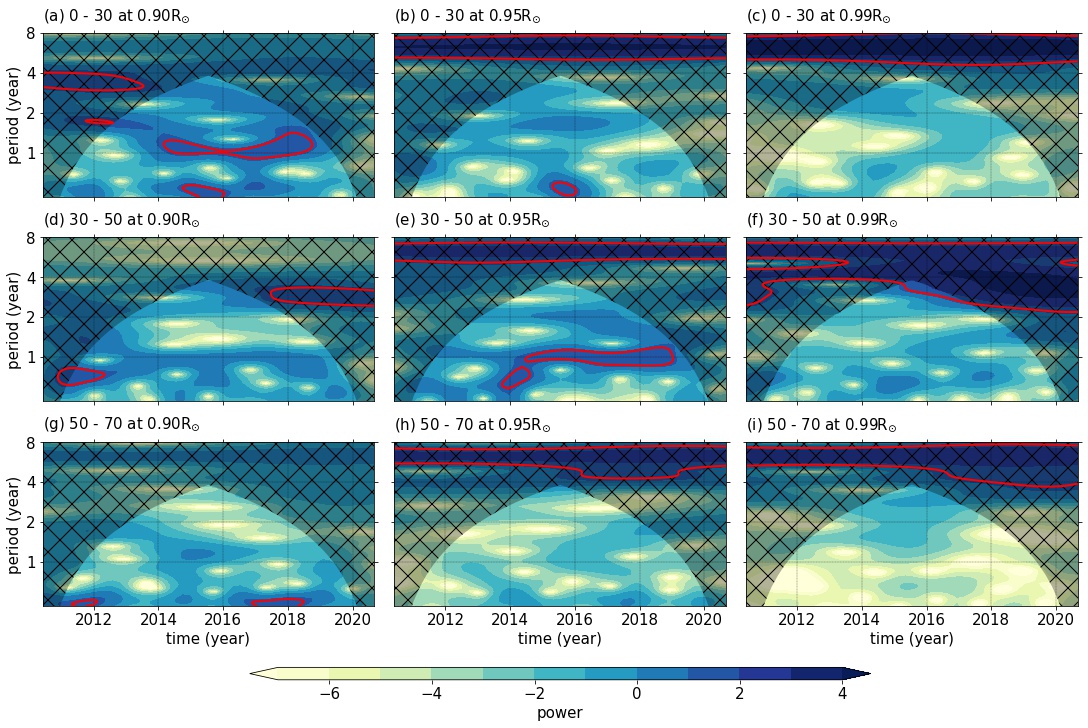}}
\caption{Same as Figure~\ref{fig:mdi_flow_CWT} but for the period covering May 2010 -- September 2020 from HMI/SDO.} 
\label{fig:hmi_flow_CWT}
\end{center}
\end{figure*}

As for solar cycle 24, the rotation rate residuals show a few QBO-like signals that are statistically significant at the 0.1 level. At the depth of 0.90$R_{\odot}$ and at the low latitudinal band the rotation rate residuals show QBO-like signals between 2014 and $\sim$2019 (the left panel of Figure~\ref{fig:hmi_flow_CWT}). At the depth of 0.95$R_{\odot}$ a QBO-like signal in the mid latitudinal band can be observed after around 2015 (the middle panel of Figure~\ref{fig:hmi_flow_CWT}).

The results generally show that both data sets display marginally significant QBO-like signals at the 0.1 level. We, therefore, decided to study the existence of the QBO-like signals further.

\subsection{The temporal evolution of the rotation rate residuals in the QBO timescale}

The common features observed in the temporal variations of the periodicities throughout solar cycles 23 and 24 are that there are strong indications for the presence of QBO-like signals, and these signals might be suppressed by the presence of longer-term ($\ge \sim$8 years) and higher amplitude variations (Figures~\ref{fig:mdi_flow_CWT} and ~\ref{fig:hmi_flow_CWT}). It is difficult to find and analyze the QBOs in data sets without any prior data preparation methods, such as filtering the signals using discrete or fast Fourier transform, spherical harmonic decomposition, passband filtering, and Empirical Mode Decomposition (EMD) analysis \citep{2014SSRv..186..359B}.

To remove the possible effects of these higher amplitude variations and to investigate the spatiotemporal behavior of the QBO-like signals even further, we decomposed our data sets using the EMD method, which decomposes the time series data into a finite number of Intrinsic Mode Functions (IMF) and they represent a set of basis functions \citep{1998RSPSA.454..903H}. This method has successfully been applied to various time-series data, including solar neutrino flux rates, solar and galactic cosmic rays measured by neutron monitors, surface magnetograms, and polar faculae data, previously \citep{2010ApJ...709L...1V, 2012ApJ...749...27V,2020MNRAS.494.4930D}.

\begin{figure*}
\begin{center}
{\includegraphics[width=5.5in]{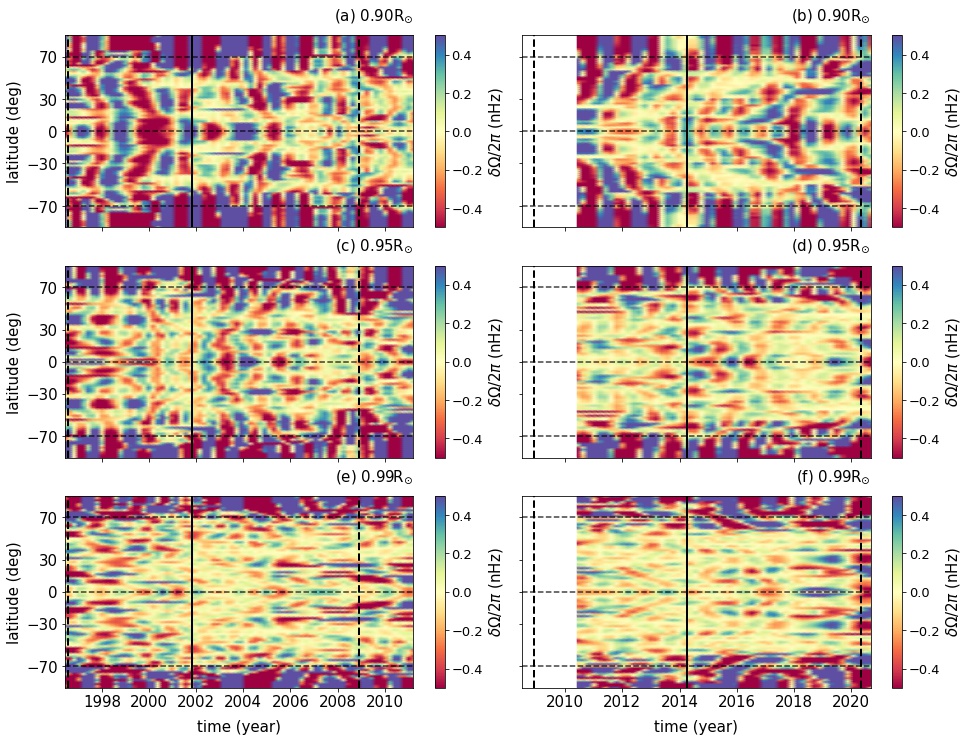}}
\caption{Same as Figure~\ref{fig:mdi_hmi_mag_flow} but for the IMF2 of the rotation-rate residuals at target depths of 0.90$R_{\odot}$, 0.95$R_{\odot}$, and 0.99$R_{\odot}$.} 
\label{fig:mdi_hmi_mag_flow_emd}
\end{center}
\end{figure*}

We decompose the rotation rate residuals at each depth and latitudinal band through solar cycles 23 and 24. It must be noted that we did not include latitudes above 70$^{\circ}$ N and S as there are some data gaps above these latitudes because of the tilt of the rotation axis of the Sun. The data is decomposed until a QBO-like oscillation with a period ranging from $\sim$1 to $\sim$4 years is found in the data. This corresponded to the second intrinsic mode function (IMF2) in the rotation rate residuals that display QBO-like signals.

The IMF2 of the rotation rate residuals during solar cycles 23 and 24 show at each depth, there are faster-than-average and slower-than-average flow bands and their amplitudes decrease as we get closer to the surface (Figure~\ref{fig:mdi_hmi_mag_flow_emd}). At the depth of 0.90$R_{\odot}$, the faster-than-average and slower-than-average flow bands form around 40$^{\circ}$ latitudes and they propagate both equator-ward and pole-ward during solar cycle 23 (Figure~\ref{fig:mdi_hmi_mag_flow_emd}a). During solar cycle 24, on the other hand, this pattern is completely different. The faster-than-average and slower-than-average flow bands form at lower latitudes and propagate pole-ward to higher latitudes (Figure~\ref{fig:mdi_hmi_mag_flow_emd}b). Another interesting feature during the rising phase of solar cycle 23 is switching between faster-than-average to slower-than-average flow bands in the higher latitudes between 50 -- 70$^{\circ}$. The switching pattern is more pronounced and clear at the 0.90$R_{\odot}$ and 0.95$R_{\odot}$, while it is more convoluted and not as pronounced at the 0.99$R_{\odot}$. This flow pattern in the higher latitudes almost vanishes during the declining phase after the maximum of the cycle at the depth of 0.99$R_{\odot}$. For solar cycle 24, however, these patches persist throughout the solar cycle (the bottom panel of Figure~\ref{fig:mdi_hmi_mag_flow_emd}). The IMF2 of the rotation rate residuals in solar cycle 23 display a different pattern at every depth than those in solar cycle 24 (Figure~\ref{fig:mdi_hmi_mag_flow_emd}).

\begin{figure*}
\begin{center}
{\includegraphics[width=5.5in]{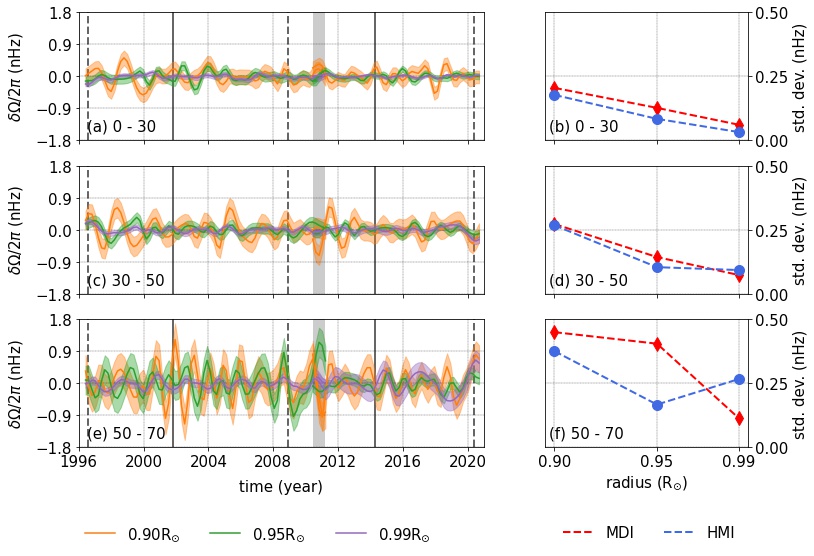}}
\caption{The left panel shows the IMF2 of the rotation-rate residuals at target depths of 0.90$R_{\odot}$ (orange), 0.95$R_{\odot}$ (green), and 0.99$R_{\odot}$ (purple) for the latitudinal bands 0 -- 30$^{\circ}$ (a), 30 -- 50$^{\circ}$ (c), and 50 -- 70 $^{\circ}$ (e) covering the period between May 1996 -- November 2008 from MDI/SOHO and the period between May 2010 -- September 2020 from HMI/SDO. The right panel shows the standard deviation in each IMF2 of the rotation-rate residuals as a function of depth for MDI (red) and HMI (blue) data. Color shaded areas show the 1${\sigma}$ standard deviation interval calculated for the IMF2 of the flow fields. The vertical solid lines show the solar cycle maxima, while the dashed lines show the cycle minima. The vertical grey shaded areas show the overlapping time period between two data sets.}
\label{fig:mdi_hmi_mag_emd_lats}
\end{center}
\end{figure*}

Following that, we averaged the IMF2 of the rotation rate residuals into low, mid, and high latitudinal bands (Figure~\ref{fig:mdi_hmi_mag_emd_lats}). The IMF2s of the rotation rate residuals calculated for each latitudinal band show QBO-like oscillations (the left panel of Figure~\ref{fig:mdi_hmi_mag_emd_lats}). The amplitude of these oscillations in each latitudinal band increases with increasing depth, except for that calculated using HMI data in 0.95$R_{\odot}$ (the right panel of Figure~\ref{fig:mdi_hmi_mag_emd_lats}). The IMF2 of the rotation rate residuals show variations with a maximum amplitude of $\sim$2 nHz during solar cycle 23 around 2002.

To investigate the spatiotemporal behavior of the QBOs-like signals in the IMF2 of the rotation rate residuals, we calculate the CWT of the data during solar cycles 23 and 24.

\begin{figure*}
\begin{center}
{\includegraphics[width=6in]{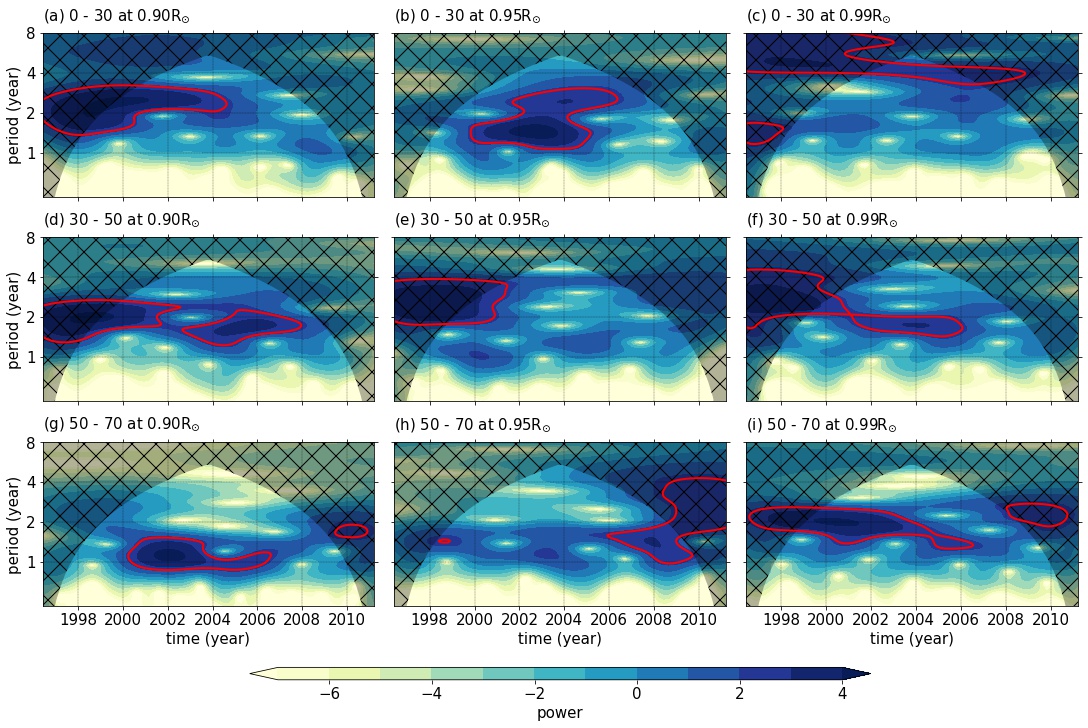}}
\caption{CWTs of the rotation rate residuals at 0.90$R_{\odot}$ (the left panel), 0.95$R_{\odot}$ (the middle panel), and 0.99$R_{\odot}$ (the right panel) for the low, mid, and high latitudinal bands for the period covering May 1996 -- November 2008 from MDI/SOHO. The red contours represent the significance level of 0.01, while the hatched area marks the cone of influence, where the CWTs might be influenced by the edge-effects.} 
\label{fig:mdi_flow_EMD_CWT}
\end{center}
\end{figure*}

There are QBO-like signals that are statistically significant at the 0.01 level at the target depth of 0.90$R_{\odot}$ at each latitudinal band. In the low latitudinal band, the QBO-like signal lasts until 2004, while it lasts throughout solar cycle 23 in the mid latitudinal band. The high latitudinal band shows a QBO-like signal, the period of which is 1.5 years between 2000 and $\sim$2004. This signal then becomes shorter in time dropping down to a period of 1 year around 2004, lasting until 2006 (the left panel of Figure~\ref{fig:mdi_flow_EMD_CWT}). However, one must be careful when evaluating periodicities close to 1 year due to the $\beta$-angle of the solar rotation axis. The target depth of 0.95$R_{\odot}$ also shows QBO-like signals significant at the 0.01 level. In the low latitudinal band, a statistically significant QBO-like signal can be observed between 2000 and 2006. In the mid-latitudinal band, a QBO-like signal is only present during the rising phase of solar cycle 23, whereas the high-latitudinal band exhibits a QBO-like signal only during the declining phase of the cycle (the middle panel of Figure~\ref{fig:mdi_flow_EMD_CWT}). At the target depth of 0.99$R_{\odot}$ at the low latitudinal band does not show any statistically significant QBO-like signals. At the middle and high latitudinal bands, on the other hand, the QBO-like signals are statistically significant at the 0.01 level and are observed between 2002 and 2006 at the mid latitudinal band, and 1998 and 2002 at the high latitudinal band (the right panel of Figure~\ref{fig:mdi_flow_EMD_CWT}).

\begin{figure*}
\begin{center}
{\includegraphics[width=6in]{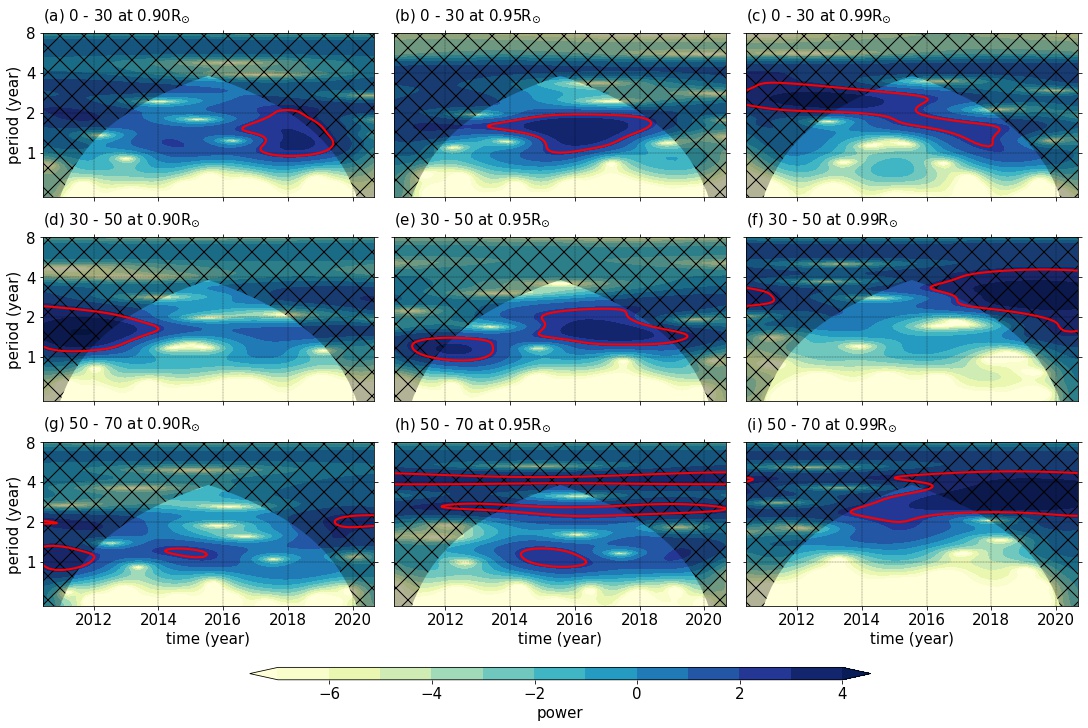}}
\caption{Same as Figure~\ref{fig:mdi_flow_EMD_CWT} but for the period covering May 2010 -- September 2020 from HMI/SDO.} 
\label{fig:hmi_flow_EMD_CWT}
\end{center}
\end{figure*}

During solar cycle 24, statistically significant QBO-like signals are observed intermittently at each depth and latitudinal band (Figure~\ref{fig:hmi_flow_EMD_CWT}). At the target depth of 0.90$R_{\odot}$ at the low latitudinal band, a statistically significant QBO-like signal is present between 2017 and 2020. At the mid latitudinal band, on the other hand, the QBO-like signal is significant until around 2014. At the high latitudinal band at the same depth, the QBO-like signal has a shorter period of around 1.2 years and lasts between 2014 and 2016 (the left panel of Figure~\ref{fig:hmi_flow_EMD_CWT}). However, similar to the signal observed for the same depth and latitudinal band in solar cycle 23, one must be careful evaluating signals that have periods close to 1 year, which might stem from the $\beta$-angle of the solar rotation axis. A statistically significant QBO-like signal at the depth of 0.95$R_{\odot}$ at the low latitudinal band extends from 2013 to 2018. The mid latitudinal band at the same depth shows QBO-like signals that are significant on the rising and declining phases of the cycle, while the high latitudinal band exhibits a QBO-like signal with a period of $\sim$2.5 years that is statistically significant throughout the cycle. Additionally, there is a statistically significant signal with a period of $\sim$1.2 year, that drops to a little below 1 year with time (the middle panel of Figure~\ref{fig:hmi_flow_EMD_CWT}). The depth of 0.99$R_{\odot}$ shows statistically significant QBO-like signals during solar cycle 24 at the low and high latitudinal bands, while the mid latitudinal band does not exhibit any significant QBO-like signal (the right panel of Figure~\ref{fig:hmi_flow_EMD_CWT}).

\section{Discussion and Conclusions} \label{sec:dis_conc}

We investigated the presence and spatiotemporal behavior of the QBO signal in the rotation rate residuals for solar cycles 23 and 24. To achieve this objective, we used data from MDI/SOHO for solar cycle 23 and HMI/SDO for solar cycle 24. We decided not to merge rotation rate residuals from the two solar observatories as using a one-year overlapping period, which might introduce systematic errors to the spliced data. It must be noted that not merging the two data did not hinder the analyses and results drawn from the data sets.

In general, even though the rotation rate residuals at each depth and latitudinal band exhibited QBO-like signals, these signals were not statistically significant at the 0.01 level. To further investigate the level of statistical significance on these results, we tried 0.05 and 0.1 significance levels, respectively. The results showed that only the 0.1 level provided some significant QBO-like signals in the data at some depths and latitudinal bands.

To further study the presence and the spatiotemporal behavior of the QBOs and remove the potential dampening effects that might be caused by variations with higher amplitude and longer periods, we decomposed the data into its intrinsic mode functions (IMFs). To achieve this objective we used the empirical mode decomposition (EMD) method following previous studies \citep{2010ApJ...709L...1V,2012ApJ...749...27V,2020MNRAS.494.4930D}. he amplitudes of the QBO-like signals found in the IMF2s of the rotation rate residuals at each latitudinal band increased with increasing depth.

The IMF2s of the rotation rate residuals at each depth and latitudinal band show QBO-like signals, which are statistically significant at 0.01 level, during solar cycles 23 and 24. The only exception is the mid latitudinal band at the target depth of 0.99$R_{\odot}$ during solar cycle 24. It must be noted that some of the periodicities observed especially in the high latitudinal bands at the target depths of 0.90$R_{\odot}$ (solar cycle 23 and 24) and 0.95$R_{\odot}$ (solar cycle 24) were in the order of $\sim$1.5 years decreasing down to 1 year as the cycles progress. These signals are close to annual signals which might be caused by the $\beta$-angle of the solar rotation axis and therefore they need to be evaluated with caution.

Our results from the IMF2s of the rotation rate residuals are in line with \citet{2012ApJ...749...27V}, who showed that the magnetic field power that is calculated using the EMD method is distributed over all latitudes and the period is concentrated around 1.5 to 4 years, which corresponds to the QBO timescales throughout solar cycles 21 and 22. The spatiotemporal behavior of the rotation rate residuals in the QBO timescale during solar cycles 23 and 24 from this study show similar patterns to those drawn from the radial and meridional components of the magnetic field measurements during solar cycles 21 and 22 \citep[the top two panels of Figure 13 in][]{2012ApJ...749...27V}. Both the IMF2s of the rotation rate residuals and the IMFs of the radial and meridional magnetic fields show pole-ward and equator-ward propagating bands as the solar cycles progress.

Using the monthly time-series of polar faculae \citet{2020MNRAS.494.4930D} reported that faculae in the high latitudes ($\ge$60$^{\circ}$) show QBO signals however these signals are decoupled showing different phase and amplitudes. Similarly, we also detected QBO-like signals in the high latitudinal band, between 50$^{\circ}$ -- 70$^{\circ}$, at the target depth of 0.99$R_{\odot}$. The high latitudinal bands in the target depths of 0.90$R_{\odot}$ and 0.95$R_{\odot}$ show QBO-like signals, however, the period of these signals are in the order of 1.5 years, which is close to the 1-year rotation signal.

Interestingly, we observed that the amplitudes of the QBO signals in the flow fields increase with increasing depth, which might point to a deeper source region for the QBO signals. In 2017, using a nonlinear shallow-water tachocline model, \citet{2017NatSR...714750D} suggested that the QBO type and Rieger-type periodicities of 150 -- 160 days can be a result of the nonlinear oscillations between Rossby waves and tachocline differential rotation, namely the tachocline nonlinear oscillations (TNOs). \citet{2017NatSR...714750D} also suggested that the TNO could potentially cause the QBO-like oscillations in solar activity indices by creating bulges and depressions in the upper boundary of the tachocline, where the toroidal magnetic field lines residing in this regime can start their buoyant rise through the convection zone when the bulging reaches its maximum upward swelling. Additionally, using a non-linear flux transport mean-field dynamo, \citet{2019A&A...625A.117I} suggested that there are indications for the QBOs to be generated via interplay between the flow and magnetic fields in the bottom of the convection zone. To achieve this result, the authors used turbulent $\alpha$-mechanism to generate the poloidal field from a pre-existing toroidal field and Lorentz-force as the saturation mechanism, which also introduces the non-linearity in the dynamo. 

We also tested our results using rotation rate residuals calculated using the optimally localized averaging method. Results from these analyses are in line with those obtained from the RLS method, further supporting the robustness of our results.

In conclusion, our results show that the QBO-like signal is dominated by higher amplitude and longer period variations in the rotation rate residuals. After removal of these effects, the QBO-like signals are present in different latitudinal bands in the rotation rate residuals at the depths of 0.90$R_{\odot}$, 0.95$R_{\odot}$, and 0.99$R_{\odot}$. Furthermore, the amplitudes of the IMF2s of the rotation rate residuals in the QBO timescales increase with increasing depth. As a follow-up to our current study, we plan to include the surface average unsigned magnetic field strengths together with flow fields to study the interactions between the magnetic and flow fields in the 11-year and the QBO timescales.

\acknowledgments

FI thanks Mausumi Dikpati and Matthias Rempel for their useful comments. RH acknowledges computing support from the National Solar Observatory. SOHO is a project of international cooperation between ESA and NASA. HMI data courtesy of NASA/SDO and the HMI science team. This work was partially funded as part of the GOES-R Series NASA-NOAA program under the University of Colorado CIRES-NOAA cooperative agreement. The views, opinions, and findings contained in this report are those of the authors and should not be construed as an official National Oceanic and Atmospheric Administration, National Aeronautics and Space Administration, or other U.S. Government position, policy, or decision. We would like to thank the reviewer for their comments, which improved the paper.

\appendix


\bibliography{paper_bibliography}{}
\bibliographystyle{aasjournal}



\end{document}